\documentclass[useAMS,usenatbib,usegraphicx]{mn2e}
\usepackage{natbib}
\bibpunct{(}{)}{;}{a}{}{,}

\usepackage{epsfig}
\usepackage{graphics}
\usepackage{float}
\usepackage{amsmath}
\usepackage{multirow}
\usepackage{longtable}
\usepackage{rotate}
\usepackage{array}
\usepackage{subfigure}
\def\phs{\phantom{$-$}}    

\def\tablecomments#1{\par\smallskip\noindent Notes. #1}
\def\plotone#1{\centerline{\psfig{figure=#1,width=\hsize,clip=}}}
\def\kms{\ifmmode{\rm km\,s^{-1}}\else\hbox{$\rm km\,s^{-1}$}\fi}
\def\nodata{\phs$\cdots$}

\let\arcdeg\degr

\setlongtables

\title[SN 2002\,ic]{On the nature of the circumstellar medium of the remarkable
       type Ia/IIn supernova, SN 2002\,ic }

\author[R. Kotak]{R. Kotak$^{1}$\thanks{E-mail: rubina@ic.ac.uk},
                                 W.P.S. Meikle$^{1}$,
                                 A. Adamson$^{2}$, S.K. Leggett$^{2}$ \\
$^{1}$ Astrophysics Group, Imperial College London, Blackett Laboratory,
               Prince Consort Road, London, SW7 2BZ, U.K. \\
$^{2}$ Joint Astronomy Centre, 660 N. A'ohoku Place, University Park, Hilo HI 96720, USA
}
                                                                                                      
\begin{document}
                                                                                                      
\date{Accepted ?? Received ??; in original form ??}
                                                                                                      
\pagerange{\pageref{firstpage}--\pageref{lastpage}} \pubyear{2004}
                                                                                                      
\maketitle
                                                                                                      
\label{firstpage}
                                                                                                      
\begin{abstract}
We present results from the first high resolution, high S/N, spectrum
of SN 2002\,ic. The resolved H$\alpha$ line has a P Cygni-type profile, 
clearly demonstrating the presence of a dense, slow-moving ($\sim$\,100\,\kms) 
outflow. We have additionally found a huge near-IR excess, hitherto unseen 
in type Ia SNe. We argue that this is due to an IR light-echo arising from 
the pre-existing dusty circumstellar medium. We deduce a CSM mass probably 
exceeding 0.3M$_{\odot}$ produced by a mass loss rate greater than several 
times $10^{-4}$~M$_{\odot}$yr$^{-1}$. For the progenitor, we favour a
single degenerate system where the companion is a post-AGB star. 
As a by-product of our optical data, we are able to provide a firm 
identification of the host galaxy of SN 2002\,ic.

\end{abstract}
                                                                                                      

\section{Introduction}
\label{sec:intro}

The lack of a convincing detection of hydrogen in the spectrum
of any type Ia supernova has been difficult to reconcile in the 
single-degenerate scenario where the non-degenerate companion,
in most candidate progenitor channels, is  hydrogen-rich. However, 
\citet{hamuy:03} recently announced the discovery of  H$\alpha$ 
emission associated with the type Ia supernova, SN 2002\,ic.
Over a time-span of about +7 to +48\,d from maximum light
\citep[$t_{B\mathrm{max}}$ = JD 2452601 = 2002 November
22;][Fig. 4]{hamuy:03}, the optical spectra of SN~2002\,ic exhibited
similar but weaker features to those of `normal' type~Ia SNe.
However, strong H$\alpha$ emission was also apparent: the H$\alpha$ 
feature consisted of a narrow component (unresolved at 300\,\kms) atop 
a broad component (FWHM $\sim$1800\,\kms). While the narrow component 
could have beeen due to an underlying HII region (but see below), 
\citet{hamuy:03} argued that the broad component arose from ejecta/circumstellar 
medium (CSM) interaction, and that this interaction also provided the 
continuum source required to dilute the spectral features of SN 2002\,ic.  
By day +48, they found that the spectrum could be equally well-matched by 
either a suitably `diluted' coeval spectrum of the type~Ia SN 1990\,N, or by 
an unmodified, roughly\footnote{Note that the explosion date of SN 1997\,cy 
is uncertain \citet{germany:00}.} 
coeval spectrum of the type~IIn SN 1997\,cy. Type~IIn supernovae (SNe~IIn) are 
so called because of the presence at early times of narrow lines in the spectra 
originating in a relatively undisturbed circumstellar medium (CSM) \citep{schlegel:90}. 
Their progenitors must therefore have undergone one or more mass-loss phases
before explosion.

In order to investigate the origin of the hydrogen emission
and hence the nature of SN~2002\,ic and its circumstellar environment,
we have acquired high resolution optical spectroscopy at +256~days,
and $HK$-band IR photometry at +278 and 380~days.  
The first results of this study are presented here.

\section{Observations}
\subsection{Optical Spectroscopy}
\label{sec:obs}
We obtained optical spectra of SN 2002\,ic and its purported host
galaxy on 2003 August 05 (=\,+256\,d) with the ESO Very Large Telescope
(VLT) Unit 2 (Kueyen) and Ultra-Violet Echelle Spectrograph (UVES).
We used a 3\arcsec\ (PA=90\arcdeg) slit which yielded a resolution of
$\sim$9\,\kms. The seeing was $\sim$\,0\farcs9. The exposure times
for SN 2002\,ic and the galaxy were 2200\,s and 1100\,s
respectively. The data were reduced in the Figaro 4
environment. Wavelength calibration was by means of a ThAr arc taken
at the end of the exposure of each of the targets. Flux calibration
was carried out with respect to the spectrophotometric standard Feige
110.

\begin{figure*}
\begin{centering}
\mbox{
\includegraphics[width=0.38\textwidth,angle=-90,clip=]{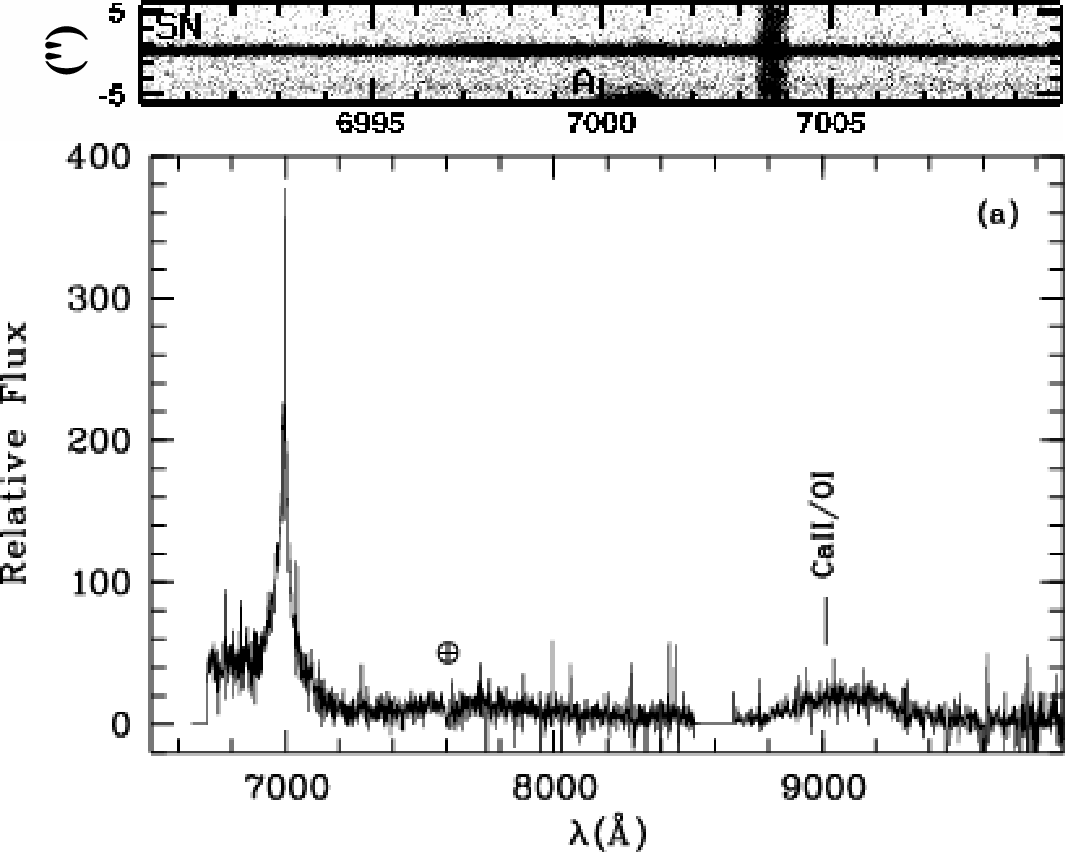}
\includegraphics[width=0.38\textwidth,angle=-90,clip=]{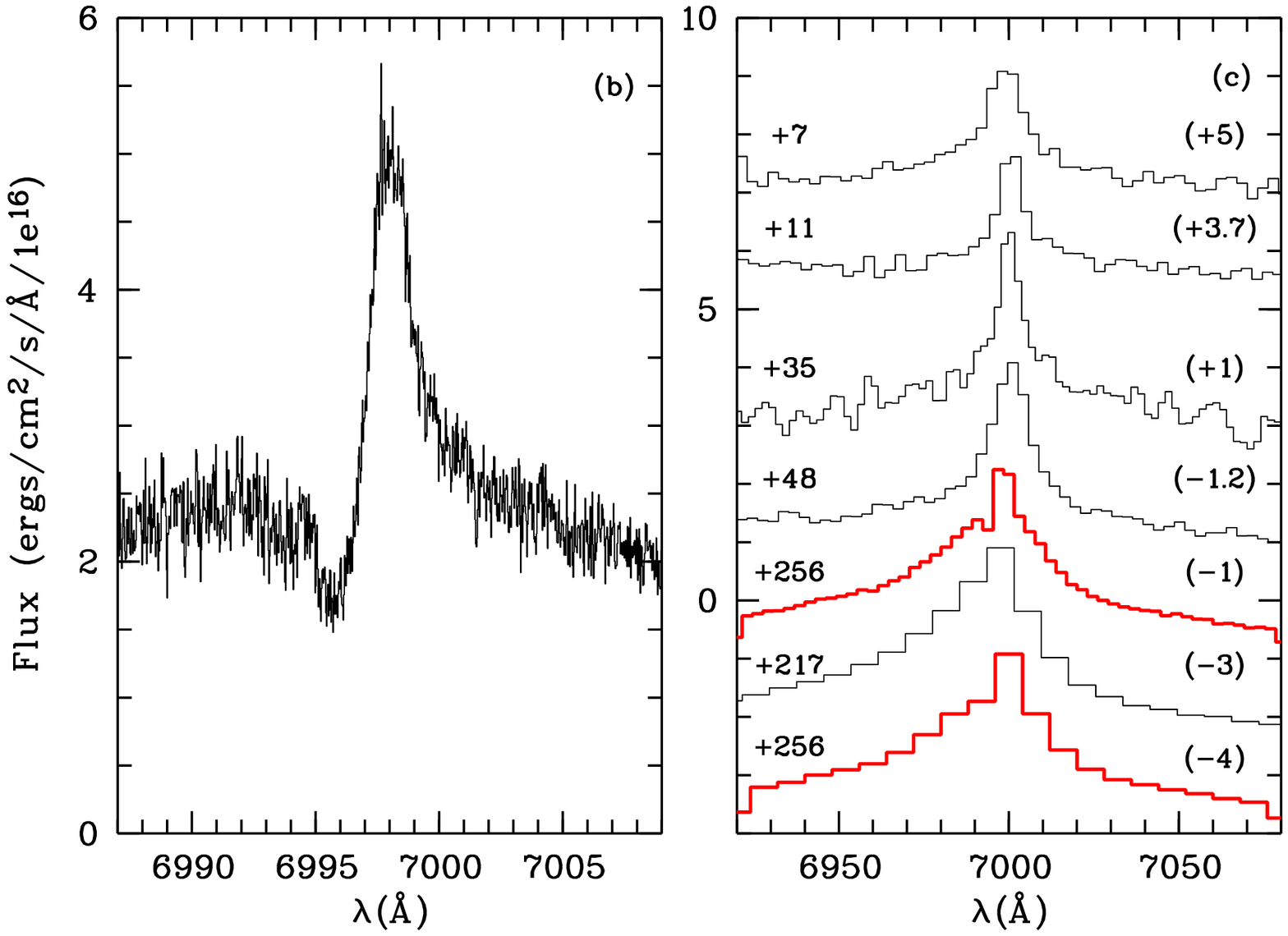}
     }
\caption{\textbf{(a):} A portion of the pipeline-reduced red spectrum
of SN 2002\,ic at +256\,d in the restframe of the supernova. The gap
at 8600\,{\AA} is due to a gap between the mosaic of the two red
chips. 
Above Fig.\,1a is shown a portion of the pre-extraction 2-D spectrum. H$\alpha$
emission from galaxy~A is marked by `A' (see also Fig. \ref{fig:chart} and 
Sec.\,\ref{sec:host}); the supernova continuum is marked by `SN'.  
\textbf{(b):} Expanded view of the unbinned H$\alpha$ profile.
\textbf{(c):} Evolution of the H$\alpha$ profile of SN 2002\,ic.  The
numbers on the left-hand side give the epoch (days) while the numbers
on the right (in brackets) give the factors by which the profiles have
been shifted vertically for clarity. The top 4 spectra are from
\citet{hamuy:03}; the 217\,d spectrum is from \citet{deng:04}; the
256\,d spectra (thick red lines) are derived from our UVES spectrum, binned
to match the resolutions of the \citet{hamuy:03} and \citet{deng:04}
spectra repectively.  }
\end{centering}
\label{fig:ha}
\end{figure*}

A portion of the UVES spectrum obtained on day~256 is shown in
Fig. \ref{fig:ha}. The spectrum is dominated by a strong H$\alpha$
feature. In addition, a weak broad feature around 9000\,\AA\ is present
which is the blend of O~I 8446\AA\ and the Ca~II IR triplet \citep{wang:04}.
In Fig.\ref{fig:ha}b we show the H$\alpha$ profile in more detail: it
comprises a narrow, but resolved, P Cygni-like profile atop a broad
emission feature.
There may also be a very broad feature present, but owing to the blue
limit of the spectrum, we are unable to give a complete description of
this feature. The \citet{deng:04} +217\,d spectrum extends further to
the blue, and they attribute the very broad feature to
[O\,I]\,6300,6364\,\AA, (FWHM\,$\sim\negthickspace\,26000$\,\kms).  We do 
not give further consideration to this component. However, we note that 
as a consequence of the presence of the very broad feature, some authors
refer to the ``broad emission feature'' mentioned above and shown in
Fig. \ref{fig:ha} as the ``intermediate component''.  We shall
continue to refer to this as the ``broad'' feature.  It is about
$5800$\,\kms\ across the base (FWHM$\sim$1550\kms).  In order to
extract more detailed information about the narrow feature, we
generated a model P~Cygni profile using a homologously expanding CSM
above a photosphere, with a rest frame wavelength equal to that of the
narrow peak and an exponentially-declining density profile. The
P~Cygni profile parameters were adjusted by eye to match the
absorption component, yielding a velocity of 100\,\kms\ at the
photosphere, and an e-folding velocity of 30\,\kms. The maximum
detectable extent of the blue wing of the absorption is
$\sim$250\,\kms. The P~Cygni model also demonstrated that the narrow
component includes additional emission not taken into account by the
absorption. Both the narrow emission component and the P~Cygni
profile suggest a CSM velocity of 80--100\,\kms. In addition, there is
a small but significant shift between the narrow emission component
and the 1500\,\kms\ component in the sense that the narrow component
peak is $111\pm7$\,\kms\ further to the red. The H$\alpha$ profile
parameters are summarised in Table~\ref{tab:ha}.

\begin{table}
\caption[]{Parameters for the H$\alpha$ profile of SN~2002ic at +256~d.}
 \begin{centering}
 \begin{tabular}{lll}
 \label{tab:gaussian}
               &  &             \\
\hline
                                      & Narrow             &  Broad \\
\hline
Peak wavelength  ({\AA})            & 6998.10$\pm$0.05   &   6995.5$\pm$0.15  \\
Width (FWHM {\AA})                  & 1.83$\pm$0.04      &   36$\pm$0.5       \\
Width (FWHM km/s)                   &  78.4$\pm$1.7      &     1500$\pm$20    \\
Intensity ($10^{-16}$\,erg\,cm$^{-2}$s$^{-1}$)           &  4.9$\pm$0.1       &   50.7$\pm$0.5     \\
\hline
\end{tabular}
\end{centering}
\label{tab:ha}
   \end{table}

In Fig. \ref{fig:ha}c we show the evolution of the H$\alpha$
profile from +7 to +256\,d. 
Our +256~day UVES spectrum provides a good match to the +217\,d
profile, showing that the narrow absorption feature which we detected
would not have been apparent in the \citet{deng:04} spectrum.  In
contrast, however, our +256\,d UVES spectrum binned to the
\citet{hamuy:03} resolution still shows evidence of the narrow
absorption feature, whereas no such feature is evident up to +48\,d.
Indeed, there is little evidence of an asymmetry in the profile at 
early times. This suggests that the photosphere is located in a
thin, opaque, cool dense shell (CDS) formed by the reverse shock 
propagating into the extended stellar envelope \citep[e.g.][]{chevfrans:94}.

\subsection{Near-infrared photometry}

We obtained a $K$-band image of SN 2002\,ic at UKIRT on 2003 August 27
(=\,+278\,d) using UIST; subsequent $H$ and $K$-band images were
obtained on 2003 December 7 (+380\,d) using UFTI. The data were
reduced using standard procedures implemented in the pipeline
software. The 278\,d $K$-band image is shown in Fig.\,\ref{fig:chart}. The
supernova is clearly visible $\sim$\,$5''$ west of galaxy~A.
\begin{figure}
\mbox{
\includegraphics[height=0.39\textwidth,clip=]{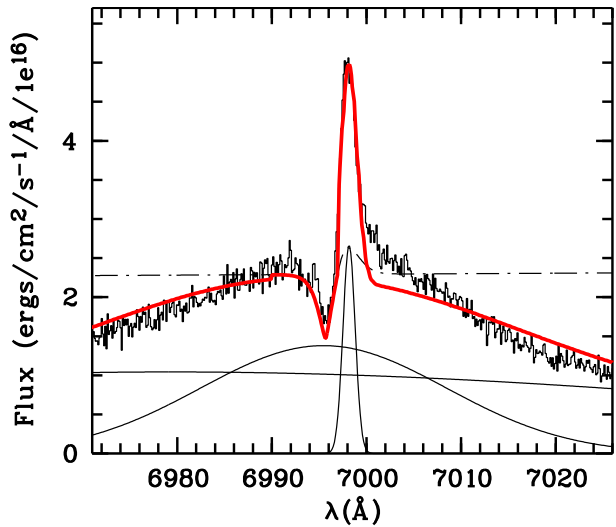}
     }
\caption{Decomposition of the H$\alpha$ profile of SN\,2002\,ic at
+256\,d. The thick line shows the combination of the three Gaussians
and the P Cygni model (dashed-dotted line).}
\label{fig:gaussfit}
\end{figure}

Magnitudes were measured using aperture photometry within the Starlink
package {\sc gaia}.
The sky background was measured using a concentric annular aperture.  
An aperture radius of $0\farcs72$ was selected, equivalent
to 6 and 8 pixels for UIST and UFTI respectively. The sky annulus was
chosen to have inner and outer radii, respectively, $\times$1.5 and
$\times$2.5 that of the aperture.  
Magnitudes were determined by comparison with the UKIRT standard star
FS\,105.  We assess the overall uncertainty in the photometry to be
about $\pm5\%$.  The $HK$ magnitudes are listed in Table \ref{tab:nirphot}.

\begin{table}
\setlength{\tabcolsep}{5pt}
\fontsize{8.}{12}\selectfont
\caption[]{Infrared magnitudes of SN~2002ic}
 \begin{centering}
 \begin{tabular}{ccccc}
 \label{tab:nirphot}
               &              &          &           &       \\ \hline
     Date      &  day   & $H$     &  $K$      &  $H-K$ \\ \hline
   20030827    & +278         &  \nodata       & +18.01$\pm0.03$    & \nodata        \\          
   20031207    & +380         & +18.98$\pm 0.02$ & +17.76$\pm0.03$    & +1.22$\pm 0.04$ \\
\hline
\end{tabular}
\tablecomments{The errors shown are statistical only.}
\end{centering}
   \end{table}

\section{Redshift of SN~2002ic and Host Galaxy Identity}
\label{sec:host}

\begin{figure}
\plotone{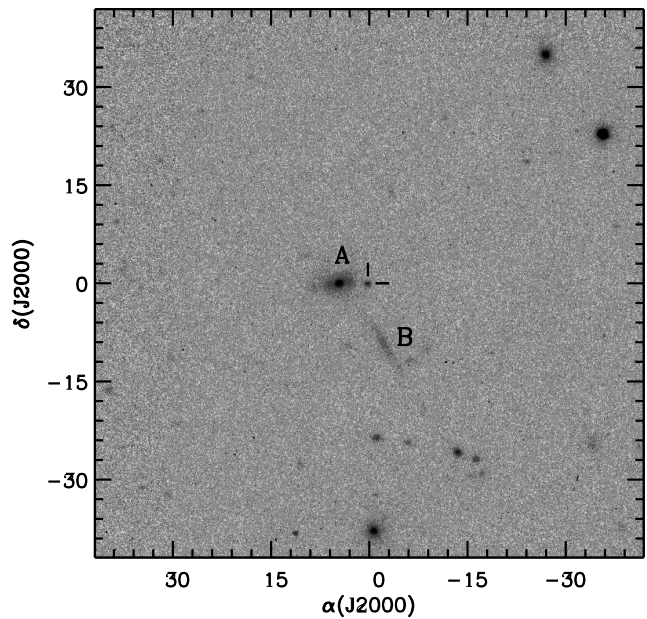}
\caption{$K$-band image of SN~2002ic obtained with UIST at UKIRT on
2003 August 27 (+278\,d). The scales give the offsets relative to SN 2002\,ic:
$\alpha_{\mathrm{J2000}}=01^h 30^m 02^s.55$;
$\delta_{\mathrm{J2000}}=+21^o 53' 06''.9$. We argue in
Sect. \ref{sec:host} that the galaxy marked `A' is the host galaxy.}
\label{fig:chart}
\end{figure}

We determined the redshift of SN 2002\,ic from the narrow emission
component of the P Cygni profile and found it to be $z=0.0663$. This
is consistent with the measurement of \citet{hamuy:03}.  Our redshift
implies a distance of $\sim$280~Mpc ($H_0$ = 70\,\kms\
Mpc$^{-1}$). According to \citet{hamuy:03}, the redshift of the galaxy
$\sim$\,5\arcsec\ E of SN 2002\,ic (marked A in Fig. \ref{fig:chart})
is 0.22 thereby ruling out association with the supernova. The only
other nearby galaxy is the one marked B in Fig. \ref{fig:chart}, lying
$\sim$\,10\arcsec\ S of the supernova; \citet{hamuy:03} suggested an
association but do not report a redshift for this galaxy. However,
our UVES spectrum of galaxy\,B indicates $z=0.0784$, i.e. it is
unlikely to be the host galaxy. Fortuitously, during our UVES
observation of the supernova, the extreme eastern end of the slit
intercepted the nuclear region of galaxy~A. We noticed an emission
feature at the corresponding spatial position (along the slit) in the
spectra, and in the same order as the H$\alpha$ feature of SN
2002\,ic, shifted only slightly in wavelength. Assuming the feature
was also due to H$\alpha$ emission (but from Galaxy~A), we derive
$z=0.0667$. This indicates that Galaxy~A must be the host of SN
2002\,ic.  Hamuy (priv. comm.)  confirms that, owing to a target
acquisition error, the redshift given for the host galaxy in
\citet{hamuy:03} is incorrect.

\section{Discussion}
\subsection{Constraints from the H$\alpha$ profile}
\begin{figure}
\plotone{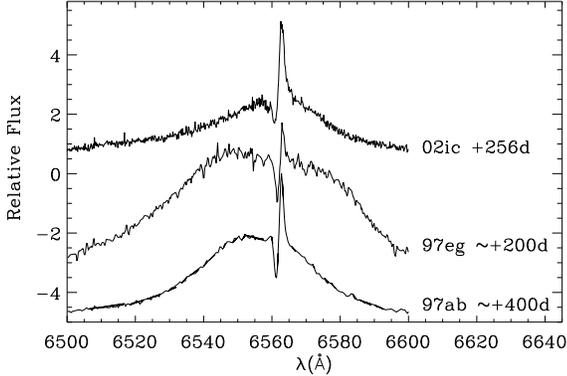}
\caption{Comparison of the H$\alpha$ profile of SN 2002\,ic with 
high-resolution spectra of two type IIn supernovae at roughly 
comparable epochs. The data for SN 1997\,ab and SN 1997\,eg are 
taken from \citet{salamanca:98,salamanca:02} and have been corrected
for their respective redshifts. } 
\label{fig:hires_sne}
\end{figure}

The narrow component of the H$\alpha$ feature suggests an origin in a
wind flowing at $v_w \sim 100$\,\kms. The presence and velocity of
the P~Cygni-like absorption immediately rules out an origin in a
line-of-sight H\,II region i.e. the narrow emission/absorption feature
is intrinsic to the supernova or its immediate environment.  
The similarity of the early-time, low-resolution spectra of SN 2002\,ic
to that of the type IIn SN 1997\,cy has been noted by several authors. 
We find that the similarity also holds at high resolution; the late-time 
H$\alpha$ profile of SN 2002\,ic is compared to those of type~IIn
supernovae observed at comparable epochs in Fig. \ref{fig:hires_sne}.
Narrow emission/absorption profiles superimposed on broad emission
features have been observed in the type~IIn events SN 1997\,ab at 425\,d
and SN 1997\,eg at +202\,d \citep{salamanca:98,salamanca:02}. For SN 1997\,ab 
the narrow absorption blue wing limit yields a velocity of 90\,\kms,
superimposed on an emission feature of $\sim$1800\,\kms\ FWHM, with
wings extending to $\sim$4000\,\kms.  For SN 1997\,eg, the
corresponding figures are 160\,\kms, 3800\,\kms and
$\sim$11000\,\kms. In both cases the narrow feature is displaced
redward of the peak of the broad component by
$\sim$\,600\,\kms. \citet{salamanca:98} attribute the apparent relative
shifts to self-absorption in the intermediate component which
preferentially attenuates the red wing. As indicated above, the
SN\,2002\,ic narrow component also exhibits a `redshift' relative to
the broad peak, but at 111\,\kms\ the shift is much smaller, suggesting
that self-absorption is less important. The luminosity of the narrow and 
intermediate H$\alpha$ features of
SN\,2002\,ic are, respectively, $5\times10^{39}$\,erg\,s$^{-1}$ and
$5\times10^{40}$~erg\,s$^{-1}$. 

The 1500\,\kms\ component is probably produced by the supernova
ejecta/wind interaction, as suggested by \citet{hamuy:03}. We note
that the width of the broad component declined from 1800\,\kms\ to
1500\,\kms\ between days +47 and +256. This gives additional weight to
the ejecta/wind interaction scenario.

The high late-time luminosity of SN~2002ic allows us to rule out
radioactive decay of $^{56}$Ni as the dominant energy source. 
The $UBVRI$ light curve \citet{deng:04} gives a luminosity of
$5\times10^{42}$~erg\,s$^{-1}$ at 250\,d, whereas the total radioactive
luminosity of 0.7M$_{\odot}$ $^{56}$Ni at this epoch is only
$1\times10^{42}$~erg\,s$^{-1}$.  Moreover, by +250\,, only about $\sim$10\% of
the decay gamma-rays would be deposited in the ejecta of the presumed
type~Ia SN ejecta \citep[e.g.][]{axelrod:80}.  Therefore, the dominant source
of energy for the broad component luminosity,
$L_{H\alpha}^{\mathrm{Broad}}$, must be due to the ejecta/CSM
interaction.  In this scenario, $L_{H\alpha}^{\mathrm{Broad}}$ is
proportional to the kinetic energy dissipation rate across the shock
front. We can use this luminosity to estimate the mass-loss rate, $\dot{M}$,
\citep{salamanca:98}: 

\begin{equation}
L_{H\alpha}^{\mathrm{Broad}} =
\frac{1}{4}\epsilon_{H_\alpha}\frac{\dot{M}}{v_{w}}v_{s}^3
\end{equation}

\noindent where $\epsilon_{H_\alpha}$ is an efficiency factor which
peaks at $\sim$0.1. $v_{s}$ is the shock velocity, and $v_w$ is the
velocity of the unshocked wind, assumed to be freely-expanding.  From
the broad H$\alpha$ line, $v_{s}$= 2900\,\kms, while the narrow
feature gives $v_{w}$=100\,\kms. Substituting into the above equation,
and using the broad component luminosity, we obtain a mass-loss rate
of $\dot{M} \sim 1.2\times
10^{-2}/(\epsilon_{H_\alpha}$/0.1)\,M$_\odot$yr$^{-1}$. 

\citet{salamanca:98} also show that by using the luminosities of both
the broad and narrow emission components, the CSM density and mass can
be estimated. Using the component luminosities given above, we find
that the number density at the CSM inner limit on day~256 is
$n_i=1.35\times10^7$\,cm$^{-3}$. Assuming that the CSM was created by
a steady wind (density $\propto$ $r^{-2}$) and that the inner limit of
the CSM on day~256 corresponds to the radius reached by the
2900\,\kms\ shock, we find a total CSM mass of
$0.037R_2/R_i$M$_{\odot}$, where $R_i$ and $R_2$ are, respectively,
the inner and outer limits of the CSM, and $R_2>>R_i$. $R_i$ can
be identified with the shock radius. 

\subsection{Constraints from NIR photometry}
We now consider the IR emission. On day +380, $H-K=1.22\pm$0.04.  Such
a colour corresponds to a 1430$\pm$40\,K blackbody.  We therefore
propose that the late-time IR emission is due to thermal emission from
hot dust associated with SN~2002\,ic or its progenitor. However, it is
possible that the IR flux contains a component due to hot
(T$\sim$10,000\,K) residual photospheric emission such as might be
produced by the shock/CSM interaction. We therefore measured the
continuum level in the vicinity of 0.94$\mu$m on day +256 and
extrapolated to day~+380 assuming an exponential decline timescale of
170\,d \citep{deng:04}. We then extrapolated to the $H$ and $K$ bands
assuming a Rayleigh-Jeans law. From this we conclude that $\sim$30\%
and $\sim$8.5\% of the $H$ and $K$ band fluxes respectively was due to
contamination by the hot photosphere.  After subtracting the
photospheric component and correcting for a Galactic extinction of
A$_V=0.198$ (NED), we find that the net IR flux on day~380 can be
reproduced by a T=1220\,K blackbody and a luminosity of
$1\times10^{42}$~erg\,s$^{-1}$.  The corresponding figure for +278\,d
(assuming the same temperature) is $0.7\times10^{42}$\,erg\,s$^{-1}$.
We note that if the IR emission were due to dust condensation in the
ejecta, for the corresponding 1220\,K blackbody to attain these
luminosities it would need to have expanded at 8000--9000\,\kms\ since
the supernova exploded.

We now consider the location and origin of the hot dust. We first note
that as with the $UBVRI$ emission, contemporary radioactive decay can
have made only a minor contribution to the late-time IR luminosity of
SN\,2002\,ic.  We also note that dust is not expected to condense in
type Ia explosions. This is consistent with the fact that to produce
the observed IR luminosity, the 1220\,K blackbody surface would have
to be located as far out as the $\sim$8500\,\kms\ region of the
ejecta. Dust condensation in such circumstances seems unlikely. We
conclude that the IR luminosity arises from a pre-existing dusty CSM.
Heating of this dust can be via (a) local heating by the ongoing
ejecta/CSM shock interaction, or (b) photon emission from the
supernova yielding an IR echo in the unshocked CSM. \citet{deng:04}
find velocities exceeding 10000\,\kms\ in C, O, and Ca, which they
attribute to the supernova ejecta.  Thus, local heating of a dusty CSM
by the ejecta/CSM shock might seem to be a possibility. However, the
the initial flash from the supernova would have evaporated any CSM
dust to at least $\sim$3500~AU for carbon-rich grains
(T$_{\mathrm{evap}}=1900~K$) and 16,500~AU for oxygen-rich grains
(T$_{\mathrm{evap}}=1500~K$) \citep{dwek:83,dwek:85}.  Yet a
10000\,\kms\ shock would have reached only 1600\,AU by +278\,d.  We
therefore rule out local heating of the dust by the ejecta/CSM shock.

We now test the possibility that the IR emission arose from the
heating of CSM dust by photons from the supernova i.e. the IR-echo
scenario. We use the bolometric light curve of \citet{deng:04}
which can be approximately described as having a peak (t=0) luminosity
of $L_0=3\times10^{43}$ erg\,s$^{-1}$ and an exponential decline timescale
of 170\,d. We attribute this slow decline to the energy released in
the interaction of the shock with the CSM. The bolometric light curve
is based on $UBVRI$ photometry, and so does not include possible
additional energy from an X-ray precursor. However, given the much
higher opacity of dust grains to UV-optical light, it is likely that
the X-ray contribution to grain heating will be small.

\citet{dwek:83} showed that the IR-echo light curve comprises an
initial plateau phase, followed by a decline.  The transition from
plateau to decline corresponds to the passing of the ellipsoid vertex
from the dust-free cavity into the region of unevaporated dust. Thus,
the radius of the dust free cavity $R_1=ct_1/2$.  We noted that the IR
flux from SN2002\,ic barely changed between days~278 and 380.  From
this we conclude that the vertex was still within the dust-free cavity
on day 380, implying a cavity radius of $R_1>32800$\,AU. We use
Eq.\,17 in \citet{dwek:83} to estimate the optical depth of dust
required to yield the observed IR flux from SN~2002\,ic. The
parameters adopted were: $R_1=32800$~AU, d = 280\,Mpc, dust IR
emissivity proportional to $\lambda^{-1}$ for wavelengths down to
0.2\,$\mu$m, a mean UV-visual absorption efficiency of 1, and an
initial dust temperature at the cavity boundary assumed to be about
equal to an evaporation temperature of 1500\,K.  The CSM was assumed
to have been produced by a steady wind so that the density is
proportional to $r^{-2}$.  We find that the IR flux at both epochs is
reproduced with a dust optical depth of $\tau_d\sim0.01$.  For a
gas-to-dust mass ratio of 160, a grain material density of 3\,gcm$^{-3}$ 
and a grain radius of 0.1\,$\mu$m, this translates \citep[from][]{dwek:83}
into a total CSM mass (including the dust-free cavity) of
$\sim$0.3$R_2/R_1$\,M$_{\odot}$, where $R_2$ is the outer limit of the
CSM and $R_2>>R_1$. Thus the CSM mass exceeds 0.3\,M$_{\odot}$. The
corresponding mass loss rate (again for $R_2>>R_1$), assuming a
wind velocity of 100\,\kms\, is $1.9\times10^{-4}$~M$_{\odot}$yr$^{-1}$ 
\citep[][Eq. 19]{dwek:83}.  It is about $\times$5 the value which \citet{dwek:83}
derived for the type~IIL SN~1979\,C.

Compared with the derivations from the H$\alpha$ line, the IR analysis
produces a lower mass-loss rate ($\sim$1\%). This discrepancy could, in
part, be due to an underestimate of the shock velocity leading to an
overestimate of the mass-loss rate derived from the broad $H{\alpha}$
line. Nevertheless, both analyses indicate a CSM mass probably
exceeding 0.3M$_{\odot}$ produced by a mass loss rate greater than
several times $10^{-4}$\,M$_{\odot}$yr$^{-1}$.
The mass-loss rate inferred from this work and that of others
is higher than expected from traditional mass loss mechanisms.
We remark that the high values of $\dot{M}$ are -- at least partly -- a 
consequence of simplifying assumptions e.g. clumped winds would mimic a high 
mass-loss rate.

The close similarity between SN 2002\,ic and type IIn SNe, has raised doubts 
as to whether SN 2002\,ic is a {\it bona fide\/} Ia event and whether other 
type IIn (i.e. core-collapse) SNe, only discovered at late epochs, may have been 
SN 2002\,ic-like events. 
We suggest that the type IIn phenomenon is predominantly
related to the amount of CSM around the progenitor system rather than
the type of explosion. 

A type 1.5 scenario i.e. the explosion of a single, massive AGB star
has been invoked as a possible progenitor of SN 2002\,ic \citep[e.g.]
[]{hamuy:03}. We point out that an extremely low metallicity would be
required to inhibit mass-loss so as to allow the degenerate core to grow
to M$_{\mathrm{Ch}}$ e.g., a 4\,M$_\odot$ star would need to have 
$\log Z/Z_{\odot} \la -3$ \citep{zijlstra:03}.
Also, a low mass-loss rate is at odds with the large
amount of CSM inferred for this event, although we cannot rule out the
possibility that the CSM is due to a binary companion.
Furthermore, SN 2002\,ic exhibited an exceptionally high maximum $V$-band 
luminosity and a much slower post-$\sim$+25days decline rate in $BVI$ than 
is seen in normal SNe~Ia. The opposite effect would be expected for type 
1.5 events i.e. the photometric and spectral evolution should be similar 
to II events at early times and dominated by the decay of radioactive Ni 
and Co at late times as for type I and IIP events \citep{ibenrenz:83}.

Any progenitor scenario must satisfy all the observational constraints
{\it viz:} the type Ia-like behaviour at early-times and the type IIn
behaviour at late-times \citep{hamuy:03}; broad profiles of Ca and O 
\citep{deng:04}; an aspherical CSM \citep{wang:04}; 
a slow-moving outflow at $\sim\negthickspace\,100$\,\kms\ and a dusty CSM. 
It must also explain the apparent rarity of SN 2002\,ic-like events.
Taking these observational constraints at face-value, we currently favour a 
system involving a post-AGB star. There are several known examples of post-AGB 
stars that have high inferred mass-loss rates and dusty discs CSM) 
\citep[e.g. IRAS 08544-4431,][]{maas:03}. 
These objects have typical outflow velocities of the order of 100\,\kms. 
Furthermore, the post-AGB phase can be relatively short, $\sim\negthickspace 10^{3}-10^{4}$
yrs. \citep{vdveen:94}. 
Further planned observations will no doubt provide more clues as to the
previous evolution of SN 2002\,ic.

\section{acknowledgements}
We thank F. Patat for expert help in setting up the observations.
Thanks also go to J. Deng, K.S. Kawabata, K. Nomoto, and Y. Ohyama for
kindly providing us with the +217\,d spectrum.
Based on DDT observations obtained with ESO Telescopes at the 
Paranal Observatories under programme ID 271.D-5021 and the United Kingdom 
Infrared Telescope which is operated by the Joint Astronomy 
Centre on behalf of the U.K. Particle Physics and Astronomy Research Council.
R.K. acknowledges support from the EC Programme `The Physics of Type Ia SNe'
(HPRN-CT-2002-00303) and interesting discussions with S. Sim and J.S. Vink.


\begin{thebibliography}{}

\bibitem[\protect\citeauthoryear{Axelrod}{1980}]{axelrod:80}
Axelrod, T.S., 1980, PhD., Univ., Santa Cruz.

\bibitem[\protect\citeauthoryear{Chevalier \& Fransson}{1994}]{chevfrans:94}
Chevalier, R.A., Fransson, C., 1994, ApJ, 420, 268

\bibitem[\protect\citeauthoryear{Deng et al.}{2004}]{deng:04}
Deng, J., Kawabata, K.S., et al., 2004, ApJ, 605L, 37

\bibitem[\protect\citeauthoryear{Dwek}{1983}]{dwek:83}
Dwek, E., 1983, ApJ, 274, 175

\bibitem[\protect\citeauthoryear{Dwek}{1985}]{dwek:85}
Dwek, E., 1985, ApJ, 297, 719

\bibitem[\protect\citeauthoryear{Germany et al.}{2000}]{germany:00}
Germany, L., Reiss, D.J., et al., 2000, ApJ, 533, 320

\bibitem[\protect\citeauthoryear{Hamuy et al.}{2003}]{hamuy:03}
Hamuy, M., Phillips, M. M., et al. 2003, Nature, 424, 651

\bibitem[\protect\citeauthoryear{Iben \& Renzini}{1983}]{ibenrenz:83}
Iben, I. Jr., \& Renzini, A. 1983, ARA\&A, 21, 271

\bibitem[\protect\citeauthoryear{Livio \& Riess}{2003}]{livioriess:03}
Livio, M., \& Riess, A., 2003, astro-ph/0308018

\bibitem[\protect\citeauthoryear{Maas et al.}{2003}]{maas:03}
Maas, T., van Winckel, H., et al., 2003, A\&A, 405, 271

\bibitem[\protect\citeauthoryear{Salamanca et al.}{1998}]{salamanca:98}
Salamanca, I., Cid-Fernandes, R., et al., 1998, MNRAS, 300, L17

\bibitem[\protect\citeauthoryear{Salamanca et al.}{2002}]{salamanca:02}
Salamanca, I., et al., 2002, MNRAS, 330, 844

\bibitem[\protect\citeauthoryear{Schlegel}{1990}]{schlegel:90}
Schlegel, E.M., 1990, MNRAS, 244, 269

\bibitem[\protect\citeauthoryear{van der Veen et al.}{1994}]{vdveen:94}
van der Veen, W., Waters, L. et al., 1994, A\&A, 285, 551

\bibitem[\protect\citeauthoryear{Wang et al.}{2004}]{wang:04}
Wang, L. Baade, D., et al., 2004, ApJ, 604L, 53

\bibitem[\protect\citeauthoryear{Ziljstra}{2003}]{zijlstra:03}
Zijlstra, A. A. 2003, MNRAS Lett., 348, 23

\end{thebibliography}
\end{document}